# Identification of ammonium salts on comet 67P/C-G surface from infrared VIRTIS/Rosetta data based on laboratory experiments. Implications and Perspectives.

Olivier Poch[1][0000-0001-6777-8296], Istiqomah Istiqomah[1], Eric Quirico[1], Pierre Beck[1,2], Bernard Schmitt[1], Patrice Theulé[3], Alexandre Faure[1], Pierre Hily-Blant[1], Lydie Bonal[1], Andrea Raponi[4], Mauro Ciarniello[4], Batiste Rousseau[1], Sandra Potin[1], Olivier Brissaud[1], Laurène Flandinet[1], Gianrico Filacchione[4], Antoine Pommerol[5], Nicolas Thomas[5], David Kappel[6,7], Vito Mennella[8], Lyuba Moroz[7], Vassilissa Vinogradoff[9], Gabriele Arnold[7], Stéphane Erard[10], Dominique Bockelée-Morvan[10], Cédric Leyrat[10], Fabrizio Capaccioni[4], Maria Cristina De Sanctis[4], Andrea Longobardo[4,11], Francesca Mancarella[12], Ernesto Palomba[4], Federico Tosi[4]

[1] Univ. Grenoble Alpes, CNRS, IPAG, 38000 Grenoble, France
[2] Institut Universitaire de France, Paris, France
[3] Aix-Marseille Université, CNRS, CNES, LAM, Marseille, France
[4] IAPS, INAF, 00133 Rome, Italy
[5] Physikalisches Institut, University of Bern, CH-3012 Bern, Switzerland
[6] Institute of Physics and Astronomy, University of Potsdam, 14476 Potsdam, Germany
[7] Institute for Planetary Research, DLR, 12489 Berlin, Germany
[8] INAF–Osservatorio Astronomico di Capodimonte, Napoli, Italy
[9] CNRS, Aix-Marseille Université, PIIM, UMR CNRS 7345, 13397 Marseille, France
[10] LESIA, Obs. de Paris, PSL, CNRS, Sorbonne Univ., Univ. de Paris, 92195 Meudon, France
[11] DIST, Università Parthenope, 80143 Napoli, Italy
[12] Dipartimento di Matematica e Fisica "E. De Giorgi," Università del Salento, Lecce, Italy
olivier.poch@univ-grenoble-alpes.fr

**Abstract.** The nucleus of comet 67P/Churyumov-Gerasimenko exhibits a broad spectral reflectance feature around 3.2 µm, which is omnipresent in all spectra of the surface, and whose attribution has remained elusive since its discovery. Based on laboratory experiments, we have shown that most of this absorption feature is due to ammonium ($NH_4^+$) salts mixed with the dark surface material. The depth of the band is compatible with semi-volatile ammonium salts being a major reservoir of nitrogen in the comet, which could dominate over refractory organic matter and volatile species. These salts may thus represent the long-sought reservoir of nitrogen in comets, possibly bringing their nitrogen-to-carbon ratio in agreement with the solar value. Moreover, the reflectance spectra of several asteroids are compatible with the presence of $NH_4^+$ salts at their surfaces. The presence of such salts, and other $NH_4^+$-bearing compounds on as-teroids, comets, and possibly in proto-stellar environments, suggests that $NH_4^+$ may be a tracer of the incorporation and transformation of nitrogen in ices, minerals and organics, at different phases of the formation of the Solar System.

**Keywords:** comets, laboratory experiments, nitrogen.

## 1   Introduction

The nucleus of comet 67P/Churyumov-Gerasimenko was mapped by the Visible, InfraRed and Thermal Imaging Spectrometer, Mapping Channel (VIRTIS-M) onboard the Rosetta spacecraft. The nucleus appeared almost spectrally uniform from 0.4 to 4 µm, characterised by a very low reflectance of few percent, a positive (red) spectral slope, and a broad absorption feature from 2.9 to 3.6 µm, centred at 3.2 µm [1]. This absorption band is the most prominent feature and is widespread on the surface of comet 67P nucleus. Earlier analyses of these reflectance spectra suggested that the darkness and slope could be due to refractory poly-aromatic organics and opaque minerals (anhydrous sulfides and Fe-Ni alloys), but the attribution of the 3.2-µm absorption feature has remained elusive [1, 2, 3]. It has been proposed to be mainly due to semi-volatile components of low molecular weight, most plausibly carboxylic (−COOH) bearing molecules or ammonium ($NH_4^+$) ions, with contribution from C−H stretches [2, 4]. To test these hypotheses, we performed a series of laboratory measurements of the reflectance spectra of carboxylic acids or ammonium salts under simulated cometary-like conditions.

## 2   Laboratory measurements

Cometary dust appears to be made of aggregates of sub-micrometre-sized (sub-µm) grains [5]. Moreover, cometary nuclei are porous (70-85 %), and their surfaces partially covered by this dust are likely to have a high porosity [6]. We have developed an experimental protocol aiming at simulating these structural properties of cometary dust, following the same methodology as described in [7].

Sub-µm grains of opaque iron sulfide (pyrrhotite, $Fe_{1-x}S$, $0 < x < 0.2$) were dispersed in a liquid water solution containing carboxylic acids or ammonium salts. The liquid mixture was then nebulised to produce droplets that were frozen in liquid nitrogen to form icy dust particles. Finally, these particles were placed inside a thermal-vacuum chamber under cometary-like temperature (170-200 K) and high vacuum ($< 10^{-5}$ mbar). Under these conditions, the water ice sublimated, and after several hours, a sublimate residue composed of a porous surface of opaque sub-µm grains mixed with a carboxylic acid or ammonium salts was obtained.

The reflectance spectra of these sublimate residues were measured from 0.4 to 4 µm using the spectro-radio-goniometer SHINE of the Cold Surfaces Spectroscopy facility of the Institut de Planétologie et d'Astrophysique de Grenoble (IPAG) [8].

## 3    Results

Figure 1 shows that the sublimate residue containing ≤ 17 wt% ammonium formate (NH$_4^+$ HCOO$^-$) exhibits an absorption band matching very closely the spectral position, asymmetric shape and maxima of absorption at 3.1 and 3.3 µm of the absorption feature observed on comet 67P.

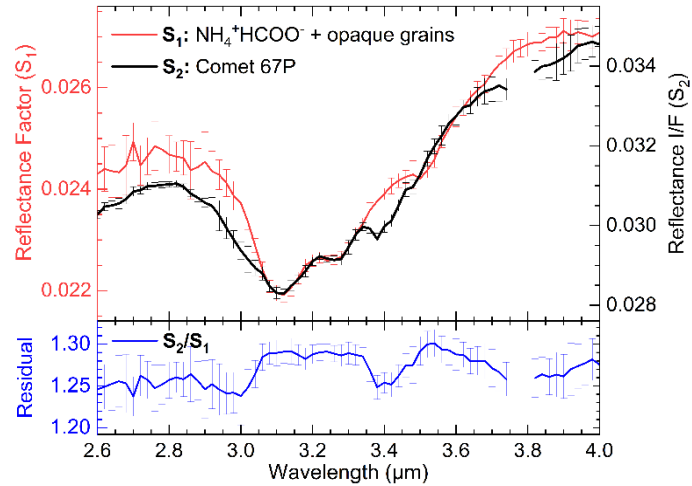

**Fig. 1.** Average reflectance spectrum of comet 67P nucleus observed by VIRTIS-M (black line, S2) [4], compared to a sublimate residue composed of ≤ 17 wt% NH$_4^+$ HCOO$^-$ and ≥ 83 wt% opaque pyrrhotite grains measured in the laboratory at 170-200 K (red line, S1). From [9]. Reprinted with permission from AAAS.

We found that other ammonium salts NH$_4^+$ X$^-$ could also provide a good match to the VIRTIS spectrum (ammonium sulfate, citrate etc.), so the exact nature of X$^-$ is unknown yet, and is certainly a mixture of various counter-ions. However, carboxylic acids, such as lactic acid, have very different absorption features [9]. Consequently, the 3.1- and 3.3-µm absorption bands observed on comet 67P appear to be due to the N−H stretching vibrations of NH$_4^+$ in ammonium salt, and several counter-ions may be expected. Additional absorption features present between 3.35 and 3.50 µm are attributed to C−H stretching vibrations of aliphatic carbonaceous compounds [4]. Other molecules have been proposed to contribute to the broad absorption, such as hydroxylated silicates [10].

The comparison of the depth of the absorption feature on the spectrum of comet 67P with spectra of mixtures of opaque grains and ammonium salts prepared and measured in the laboratory, suggests that the dust may contain several percent of ammonium salts. The abundance of ammonium salts is unknown, but could be from few percent up to 40 wt% in the cometary dust. The determination of a more precise concentration of the salts remains difficult because the band depth depends on many physical properties of the surface (grain sizes, mixing mode etc.) which may differ in our experiments compared to the comet. Moreover, one cannot exclude that the concentration of $NH_4^+$ at the surface may be different from at depth [11].

## 4 Conclusions and Perspectives

### 4.1 Independent identifications of $NH_4^+$ salts by Rosetta instruments

The surface spectrum of comet 67P shown in Figure 1 is an average of all the dataset acquired at the beginning of the mission, from August to September 2014 [4, 9]. When the comet was closer than 2.25 AU from the Sun (around perihelion, so between April 2015 and January 2016), some dust particles from the coma entered the ROSINA-DFMS mass spectrometer, where they were heated at 273 K [12,13]. All fragments ($NH_3$, $H_2S$, HCl, etc.) typical of the thermal decomposition of several ammonium salts ($NH_4^+$ $SH^-$, $NH_4^+$ $Cl^-$, $NH_4^+$ $F^-$, $NH_4^+$ $CN^-$, $NH_4^+$ $OCN^-$, $NH_4^+$ $HCOO^-$, $NH_4^+$ $CH_3COO^-$) were detected by the mass spectrometer in the hours following the entry of the dust particles ([12,13], see Rubin *et al.* in this volume for a presentation of ROSINA results). VIRTIS-M and ROSINA-DFMS instruments have thus independently identified the presence of ammonium salts in the cometary dust [9, 12]. The dust analyzer instrument onboard Rosetta, COSIMA, has not reported the detection of ammonium salts, most probably because these relatively volatile compounds have completely sublimated during the long pre-analysis storage of the dust particles at 283 K inside this instrument [14].

### 4.2 A new cometary reservoir of nitrogen

The 3.2-μm absorption feature observed on the entire surface of the comet 67P nucleus can be mainly attributed to ammonium salts and aliphatic organic matter. The exact concentration of the salts remains unknown. If they comprise more than about 5 wt% of the dust, these salts could be the major reservoir of nitrogen in the comet, dominating over the refractory organic matter and volatile species. Consequently, the abundance of nitrogen in comet 67P may be closer to the solar elemental composition than previously thought [9].

Moreover, $NH_4^+$ salts are possibly present not only on comet 67P but on a multitude of comets [12]. Indeed, the observations of higher $NH_3/H_2O$ ratios in comae with decreasing perihelion distances (5% for D/2012 S1 (ISON) at 0.46 AU compared to 0.2% for 67P at 2 AU) [15, 12] may be explained by the relatively high sublimation temperature of $NH_4^+$ salts, decomposing more efficiently in $NH_3$ when closer to the Sun [9, 12, 16]. A significant fraction of nitrogen may thus be locked in ammonium salts in a multitude of comets.

### 4.3 $NH_4^+$ on other comets, asteroids, and in proto-stellar environments?

As mentioned earlier [17], the absorption feature of comet 67P shares similarities with other features observed around 3.1 μm on several asteroids [9]. Could some of these absorption features be partly due to ammonium salts? Ammoniated phyllosilicates and ammonium salts have been detected on the dwarf planet Ceres ([18], see also Raponi *et al.* in this volume). Absorption bands similar of $NH_4^+$-bearing phyllosilicates [19] have also been detected on dust grains returned from the asteroid Ryugu by the Hayabusa2 mission ([20], see also Brucato *et al.* in this volume for a presentation of this mission). Could Ceres and Ryugu have inherited ammonium ions from outer Solar System objects similar to comet 67P? In addition, ammonium salts have been tentatively detected in the icy grains around proto-stars [21]. $NH_4^+$-bearing compounds may thus be tracers of the incorporation and transformations of nitrogen, in ices, minerals and organics, from the proto-solar nebula to comets and asteroids.

Future spectroscopic observations of such environments and bodies over an extended wavelength range up to the mid-infrared, as permitted with the James Webb Space Telescope, may provide new insights on the presence and abundance of $NH_4^+$, and the nature of its counter-ions, in these various environments and objects.

### 4.4 Ammonium salts formation and origins

Laboratory experiments have shown that ammonium salts can be produced in/on solid ices via purely thermal reactions [22, 23, 24],

- by acid-base reactions from 10-30 K to higher temperatures:
  $NH_3 + HCOOH \rightarrow NH_4^+ + HCOO^-$
- or nucleophilic addition of $NH_3$ with $CO_2$:
  $2NH_3 + CO_2 \rightarrow NH_4^+ + NH_2COO^-$ .

The presence of surfaces outside and inside the ice (e.g. porous dust surfaces, and/or surfaces formed by thermal processing of ices: segregation, crystallization or sublimation) enables diffusion of the reactants and enhance the formation of these salts [22, 24]. So $NH_4^+$ salts can start to form early in pre- or proto-stellar ices (at 10-30 K), but their formation is probably more efficient as the temperature increases at later stages, in protoplanetary disks and in comets approaching the Sun.

### 4.5 Potential influences of $NH_4^+$ salts on planetary formation processes

Understanding the origins and evolutions of these salts may help to know how nitrogen was incorporated and distributed from interstellar environments to planetary systems. Ammonium salts may have played important roles at different stages of this evolution:

- Once being formed, they may influence the chemical evolution of icy grains mantles (reactivity, phase changes, etc.).
- If abundant enough, semi-volatile ammonium salts could enhance the stickiness of icy/dusty grains, contributing to accretion [25, 26]. Moreover, some of these nitrogen-bearing salts are solid up to more than 200 K, so they may provide nitrogen closer to the Sun than the $N_2$ and $NH_3$ snow lines [27].

- Salts lower the melting point of water in the subsurface of icy bodies (icy moons of the giant planets, Ceres), with implications for cryo-volcanism and habitability (pH, water activity) [28].
- Finally, $NH_4^+$ salts are known to facilitate several prebiotic reactions occurring in liquid water, such as the phosphorylation of RNA [29].

## 5      Acknowledgements

O.P. acknowledges a post-doctoral fellowship from Centre National d'Etudes Spatiales (CNES). This work was supported by the CNES, the Agence Nationale de la Recherche (Classy grant, ANR-17-CE31-0004), the European Research Council (SOLARYS grant, ERC-CoG2017-771691), the Italian Space Agency and the DLR. This work takes advantage of the collaboration of the ISSI international team "Comet 67P/Churyumov-Gerasimenko Surface Composition as a Playground for Radiative Transfer Modeling and Laboratory Measurements", number 397. D.K. acknowledges DFG-grant KA 3757/2-1.